\documentclass[12pt]{article}
\usepackage{graphics}
\usepackage{epsfig}
\usepackage{subfigure}
\usepackage{graphics}
\usepackage{amsmath}
\usepackage{amsfonts}
\usepackage{amssymb}
\usepackage{url}
\usepackage{hyperref}
\begin{document}
%%%%%%%%%%%%%%%%%%%%%%%%%%%%
\begin{titlepage}
%%%%% PREPRINT NUMBERS %%%%%%

%%%%%%%%%%%%%%%%%%%%%%%%%%%%%%
\vspace{2\baselineskip}
%%%%%%%%%%%%%%%%%%% TITLE %%%%%%%%%%%%%%%%%%
\begin{center}
{\Large\bf 
Field limit and nano-scale surface topography of superconducting radio-frequency cavity made of extreme type II superconductor
}
\end{center}
%%%%%%%%%%%%%%%% AUTHORS %%%%%%%%%%%%%%%%%%%%%%%
\vspace{0.2cm}
\begin{center}
{\large
Takayuki Kubo%$^{\,\,a}$
%\footnote{\tt E-mail:kubotaka@post.kek.jp}
}
\end{center}
%%%%%%%%%%%%%%%%%%%%%%% AFFILIATION %%%%%%%%%%%%
\vspace{0.2cm}
\begin{center}
%${}^{a}$ 
{KEK, High Energy Accelerator Research Organization, 1-1 Oho, Tsukuba, Ibaraki 305-0801, Japan}
\end{center}
%\vskip 5mm

\begin{abstract}%
The field limit of superconducting radio-frequency cavity made of type II superconductor with a large Ginzburg-Landau parameter is studied with taking effects of nano-scale surface topography into account. 
If the surface is ideally flat, the field limit is imposed by the superheating field. 
On the surface of cavity, however, nano-defects almost continuously distribute and suppress the superheating field everywhere. 
The field limit is imposed by an effective superheating field given by the product of the superheating field for ideal flat surface and a suppression factor that contains effects of nano-defects. 
A nano-defect is modeled by a triangular groove with a depth smaller than the penetration depth. 
An analytical formula for the suppression factor of bulk and multilayer superconductors are derived in the framework of the London theory. 
As an immediate application, the suppression factor of the dirty Nb processed by the electropolishing is evaluated by using results of surface topographic study. 
The estimated field limit is consistent with the present record field of nitrogen-doped Nb cavities. 
For a further improvement of field, a surface processing technology that can realize a surface with a smaller slope-angle distribution is necessary. 
Suppression factors of surfaces of other bulk and multilayer superconductors, and those after various surface processing technologies can also be evaluated by using the formula. 
\end{abstract}

\end{titlepage}

%\subjectindex{superconducting radio-frequency cavity, field limit}

%\maketitle

%\tableofcontents

%%%%%%%%%%%%%%%%%%%%%%
%%%%%%%%%%%%%%%%%%%%%%
\section{Introduction}
%%%%%%%%%%%%%%%%%%%%%%
%%%%%%%%%%%%%%%%%%%%%%

The superconducting (SC) radio-frequency (RF) cavity is a key component of modern particle accelerators~\cite{hasan}. 
Its performance is described by the peak surface magnetic-field, $B_{\rm pk}$, and the quality factor, $Q_0$. 
$B_{\rm pk}$ is proportional to the accelerating gradient defined by the average electric field that the charged particles see during transit, 
which determines necessary accelerator length to achive a target particle energy. 
$Q_0$ is defined by the ratio of stored energy to dissipation per RF cycle.  
A higher $Q_0$ is necessary to maintain the cryogenic load manageable as $B_{\rm pk}$ increases.
Improvements of both $B_{\rm pk}$ and $Q_0$ are vital technological challenges for a future high-energy accelerator, 
such as the International Linear Collider $1\,{\rm TeV}$-upgrades~\cite{tdr}.

The recently-developed surface processing recipe involving impurity-doping 
enabled to obtain higher $Q_0$ than what had been achieved previously~\cite{grassellino, dhakal_IPAC12, dhakal}.
However, typical achievable $B_{\rm pk}$ of impurity-doped Nb cavities remain rather small~\cite{crawford, romanenko, gonnellaLINAC14, furutaLCWS14, gengLCWS14, dhakal_ieee, dhakalIPAC14, ciovati}. 
The record value of $B_{\rm pk}$ of nitrogen-doped Nb cavities is $1.3\times 10^2\,{\rm mT}$, 
and that of titanium-alloyed Nb cavities is $1.2\times 10^2 \,{\rm mT}$. 
The multilayer coating~\cite{gurevich, kubo} also attracts attention as an idea for realizing high-field and high-$Q_0$ SCRF cavity, but still is in a proof-of-concept stage~\cite{antoine, roach2013}. 
How large $B_{\rm pk}$ can be achieved by cavities based on these new technologies is a topic of interest in the SCRF community.

The fundamental limit of $B_{\rm pk}$ is thought to be imposed by the superheating field, $B_{s}$, at which the Bean-Livingston (BL) barrier for penetration of vortices disappears~\cite{bean, gurevich_review, gurevich_ciovati}. 
For a type II SC with a large Ginzburg-Landau (GL) parameter, $B_s$ is computed in all temperature range below the critical temperature $T_c$~\cite{galaiko, catelani, lin}, 
which is applicable to materials like the dirty Nb, ${\rm Nb_3 Sn}$, ${\rm NbN}$ etc., 
if the surface can be regarded as ideally flat. 
According to studies on surface topographies of SCRF materials~\cite{pehlivan, xu, roach}, however,  
the surface is covered by multi-scale structures characterized by the fractal nature~\cite{avnir, takayasu}. 
In particular, nano-scale defects distribute with much higher density than micrometer- or millimeter-scale defects 
and almost continuously exists on the surface. 
$B_{s}$ is reduced at each nano-defect. 
Then the limit of $B_{\rm pk}$ of a real cavity would be imposed not by $B_s$ but by an effective superheating field $\widetilde{B}_s=\eta B_{s}$, 
where $\eta$ is a suppression factor that contains effects of nano-defects.

In this paper, the field limit of SCRF cavity made of a type II SC with a large GL parameter is studied with taking effects of nano-defects into account. 
We consider a simple model of nano-defect and derive a formula for suppression factor, $\eta$, in the framework of the London theory. 
Combining the formula with data of surface topographic studies, 
$\widetilde{B}_s$ of materials with large GL parameters can be evaluated.

%%%%%%%%%%%%%%%%%%%%%%
%%%%%%%%%%%%%%%%%%%%%%
\section{Model and calculations of suppression factor}
%%%%%%%%%%%%%%%%%%%%%%
%%%%%%%%%%%%%%%%%%%%%%

%%%%%%%%%%%%%%%%%%%%%%%%%%%%%%%%%%%%%%%%%%%
\subsection{Model}
%%%%%%%%%%%%%%%%%%%%%%%%%%%%%%%%%%%%%%%%%%%

\begin{figure}[*t]
   \begin{center}
   \includegraphics[width=0.5\linewidth]{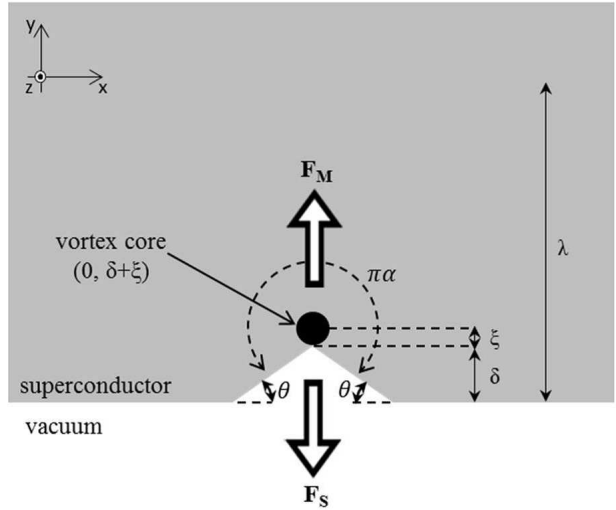}
   \end{center}\vspace{-0.2cm}
   \caption{
Triangular groove with a depth smaller than the penetration depth as a model of nano-defect.  
   }\label{fig1}
\end{figure}

There exist several types of defect models that treat the suppression of $B_s$~\cite{buzdin2, vodolazov, bass,  buzdin, aladyshkin, clem}. 
In particular, Buzdin and Daumens~\cite{buzdin} and Aladyshkin et al.~\cite{aladyshkin} studied the groove with triangular section and derived simple formulae for locally suppressed $B_s$, 
which can incorporate a geometry of defect via an angle parameter 
and are useful for modeling surface topographies~\cite{dzyuba}.   
Their formulae are, however, derived under an assumption that the groove has an infinite depth, 
which can be applied to a defect with a depth much larger than penetration depth ($>\mu{\rm m}$)~\cite{dzyuba}, 
but can not be applied to that smaller than penetration depth ($<\mathcal{O}(10^2)\,{\rm nm}$). 
We consider a model of a groove with triangular section as shown in Fig.~\ref{fig1}. 
Gray and white regions represent an SC and the vacuum, respectively. 
The surface of SC is parallel to the $xz$ plane. 
The groove and the applied magnetic-field are parallel to the $z$-axis. 
A geometry of groove is specified by a depth, $\delta$, and an angle, $\pi \alpha$ ($1<\alpha<2$). 
A slope angle is then given by $\theta = \pi (\alpha-1)/2$. 
The SC material is a type II SC with a large GL parameter, 
and its coherence length and penetration depth are given by $\xi$ and $\lambda\,(\gg \xi)$, respectively. 
Furthermore, the assumption $\xi \ll \delta$ is necessary for treating the model in the framework of the London theory. 
The parameters of the model are summarized in Table~\ref{table1}.

\begin{table}[tb]
\caption{Parameters of the model. }%%%Table caption goes here
\label{table1}
\centering
\begin{tabular}{lcc}%%%The number of columns has to be defined here
\hline
\hline
Parameter         & Character & Assumption and range \\ %& Parameter \\ %%%% Table body
\hline
Coherence length  & $\xi$     &             \\
Penetration depth & $\lambda$ & $\lambda \gg \xi$ \\
Depth of groove      & $\delta$  & $\xi \ll \delta \ll \lambda$ \\
Angle of groove     & $\alpha$  & $1 < \alpha < 2$ \\
\hline
\hline
\end{tabular}
\end{table}%%%End of the table

$\widetilde{B}_s$ of this model can be evaluated by calculating forces acting on a vortex in the framework of the London theory~\cite{bean, kuboSRF2013, kuboIPAC14}.  
Suppose there exist a single vortex next to the groove, as shown in Fig.~\ref{fig1}. 
This vortex feels two distinct forces, ${\bf F}_{\rm M}$ and ${\bf F}_{\rm S}$, 
where ${\bf F}_{\rm M}$ is a force from an external magnetic-field, and 
${\bf F}_{\rm S}$ is that from the surface. 
The former draws the vortex into the inside, 
and the latter pushes the vortex to the outside. 
When the total force ${\bf F}_{\rm tot} ={\bf F}_{\rm M} + {\bf F}_{\rm S}$ vanishes, the derivative of free-energy with respect to vortex position vanishes, namely, the BL barrier disappears.  
Thus $\widetilde{B}_s$ is a field at which these two competing forces are balanced. 
In order to evaluate these forces, 
current densities at the vortex position are necessary, 
which can be calculated by using a powerful mathematical tool; the method of conformal mapping. 
A lot of examples of the technique relevant to this work are summarized in a text book~\cite{laura} or a previous study~\cite{clem}.

It should be noted that calculations based on the London theory suffers divergences of current density at the vortex core and the sharp corner. 
These artifacts disappear if the suppression of superfluid density by current and the non-locality of current-field relation of the Bardeen-Cooper-Schrieffer (BCS) theory are taken into account. 
We choose instead to introduce a cutoff scale of the London theory, $\xi$~\cite{bean, kuboSRF2013, kuboIPAC14}. 
This small-scale cutoff makes an effective minimum distance between the surface and an axis of vortex core, 
and ${\bf F}_{\rm S}$ becomes finite. 
The screening-current density at the sharp corner diverges, 
but that at a vortex next to the corner becomes finite due to an effective minimum distance between them, 
and ${\bf F}_{\rm M}$ also becomes finite. 
Note that introducing a finite curvature-radius of the corner~\cite{laura, clem} makes the model more realistic, eliminates a divergence of screening-current density, and makes ${\bf F}_{\rm M}$ finite without cutting off, 
which might be a work to be addressed in a future, 
but ${\bf F}_{\rm S}$ diverges if a cutoff is not introduced. 
A cutoff is indispensable as long as $\widetilde{B}_s$ is evaluated in the framework of the London theory.

The model introduced above and following calculations based on the London theory only gives qualitative results, 
but may be a good starting point to evaluate local reduction of the surface barrier by small topographic defects and the maximum field at which this surface barrier vanishes.

%%%%%%%%%%%%%%%%%%%%%%%%%%%%%%%%%%%%%%%%%%%
\subsection{Force from an external magnetic-field}
%%%%%%%%%%%%%%%%%%%%%%%%%%%%%%%%%%%%%%%%%%%

%
\begin{figure}[*t]
   \begin{center}
   \includegraphics[width=0.5\linewidth]{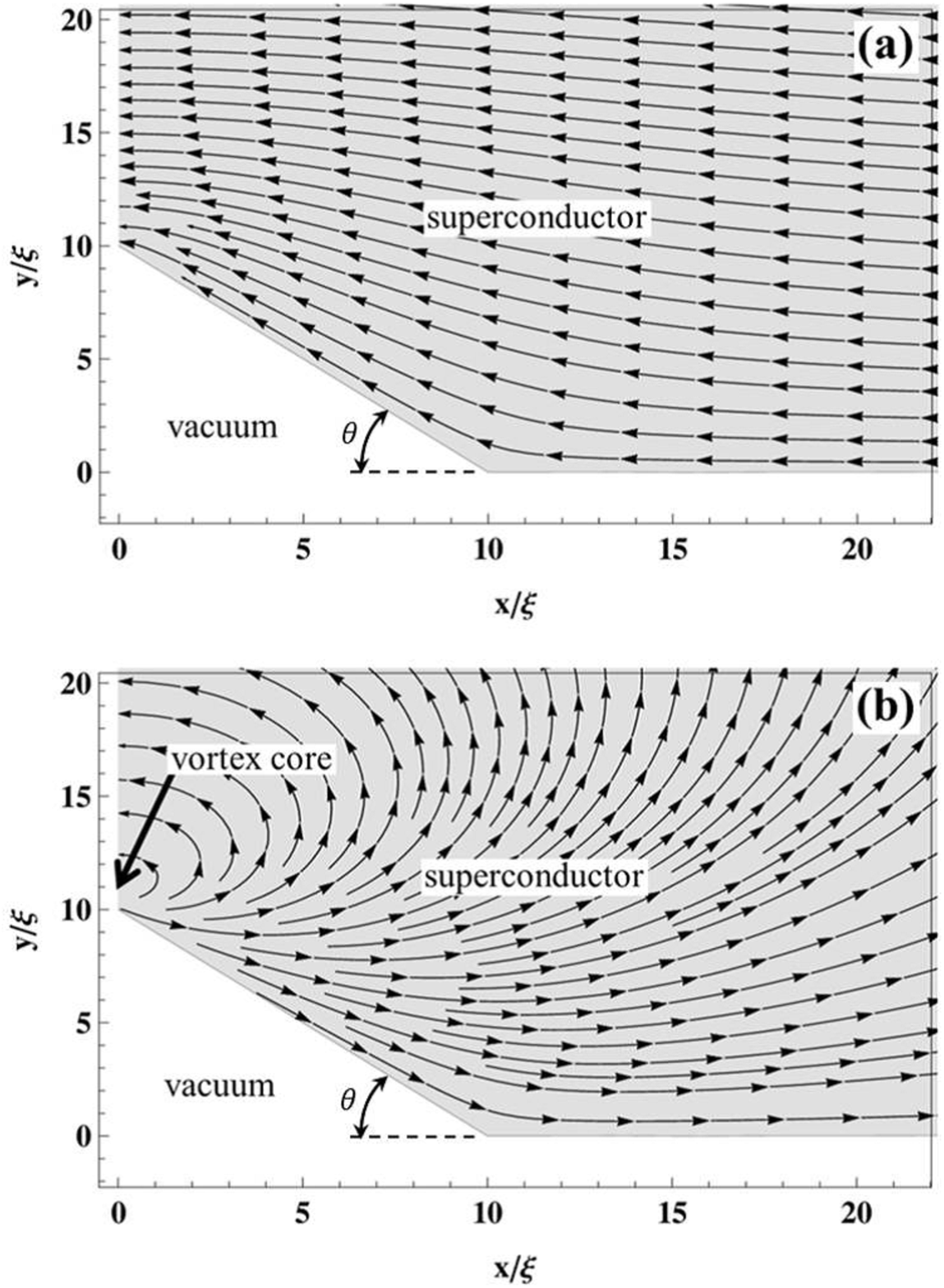}
   \end{center}\vspace{-0.2cm}
   \caption{
(a) ${\bf J}_{\rm M}$ and (b) ${\bf J}_{\rm V+I}$ calculated from Eq.~(\ref{eq:JM}) and Eq.~(\ref{eq:JVI}), respectively. 
An arrow represents a direction of current at each point.  
A depth and an angle are assumed to be $\delta=10\xi$ and $\pi\alpha=3\pi/2$ ($\theta =45^{\circ}$), respectively. 
   }\label{fig2}
\end{figure}

The force from an external magnetic-field, ${\bf F}_{\rm M}$, can be derived from the formula~\cite{tinkham}, 
${\bf F}_{\rm M}={\bf J}_{\rm M} \times \phi_0 \hat{{\bf z}}$, 
where ${\bf J}_{\rm M}$ is the screening-current, $\phi_0=2.07\times10^{-15}\,{\rm Wb}$ is the flux quantum, and $\hat{\bf z}$ is the unit vector parallel to the $z$-axis.   
The calculation of ${\bf J}_{\rm M}$ is a two-dimensional problem that can be formulated on the complex plane of the complex variable $\zeta=x+iy$,  
which can be easily solved by using the conformal mapping that maps the SC with flat surface on the complex $w$-plane into that with groove on the $\zeta$-plane (see Appendix~\ref{appendix_JM}).  
Then the components of ${\bf J}_{\rm M}$ are given by~\cite{kuboLINAC14} 
\begin{eqnarray}
J_{{\rm M}x}(x,y) - i J_{{\rm M}y}(x,y) 
= -\frac{J_0}{f(w)}\biggr|_{w=F^{-1}(\zeta)}    \,,
\label{eq:JM}
\end{eqnarray}
where $J_0$ is a screening current far from the groove,
$F^{-1}$ is the inverse of the map $F(w) = K_1 \int_0^w \!\! f(w')  dw' + K_2 $, 
$f(w) = w^{\alpha -1}(w^2 - 1)^{-\frac{\alpha-1}{2}}$, 
$K_1= \sqrt{\pi} \delta/[\Gamma(\frac{\alpha}{2}) \Gamma(\frac{3-\alpha}{2}) \sin\frac{\pi(\alpha-1)}{2}]$, 
and $K_2 = i\delta$ (see Appendix~\ref{appendix_K1K2}). 
The screening current distribution calculated from Eq.~(\ref{eq:JM}) is shown in Fig.~\ref{fig2}(a).  
Then ${\bf J}_{\rm M}$ at the vortex position $(x,y)=(0,\delta +\xi)$ is given by
$J_{{\rm M}x}(0, \delta+\xi)  - i J_{{\rm M}y}(0, \delta+\xi)  = -J_0/f(i\epsilon)= - ( K_1/\alpha \xi )^{\!\frac{\alpha-1}{\alpha}} \!J_0$, 
where $i\epsilon \equiv F^{-1}(i (\delta+\xi))$, 
$\epsilon = (\alpha \xi/K_1)^{\frac{1}{\alpha}}+ \mathcal{O}(\epsilon^2)$, 
and the term $\mathcal{O}(\epsilon^2)$ is negligible as long as our assumptions in Table~\ref{table1} are satisfied. 
Then we obtain 
\begin{eqnarray}
{\bf F}_{\rm M}(0, \delta+\xi) 
= \biggl( \frac{\sqrt{\pi}}{\Gamma(\frac{\alpha}{2}) \Gamma(\frac{3-\alpha}{2}) \alpha  \sin\frac{\pi(\alpha-1)}{2}} \frac{\delta}{\xi}  \biggr)^{\!\!\!\frac{\alpha-1}{\alpha}} 
\!\!\phi_0 J_0 \, \hat{\bf y} ,
\label{eq:FM}
\end{eqnarray}
where $\hat{\bf y}$ is the unit vector parallel to the $y$-axis. 
Note that, when $\alpha \to 1$, Eq.~(\ref{eq:FM}) reproduces the force acting on a vortex near a flat surface, $F_{\rm M0}\equiv \phi_0 J_0$. 
In Fig.~\ref{fig3}, $F_{\rm M}$ in units of $F_{\rm M0}$ are shown as functions of $\pi \alpha$. 
Larger $\alpha$ and $\delta$ induce an larger enhancement of $F_{\rm M}$.  
This behavior can be understood from a current flow:  
as $\alpha$ increases, a flow becomes rapidly bent, and as $\delta$ increases, a volume of flows affected by the groove increases.  
Then $J_{\rm M}$ and thus $F_{\rm M}$, which is proportional to $J_{\rm M}$, are enhanced as $\alpha$ and $\delta$ increase.

%%%%%%%%%%%%%%%%%%%%%%%%%%%%%%%%%%%%%%%%%%%
\subsection{Force from the surface}
%%%%%%%%%%%%%%%%%%%%%%%%%%%%%%%%%%%%%%%%%%%

%
\begin{figure}[*t]
   \begin{center}
   \includegraphics[width=0.5\linewidth]{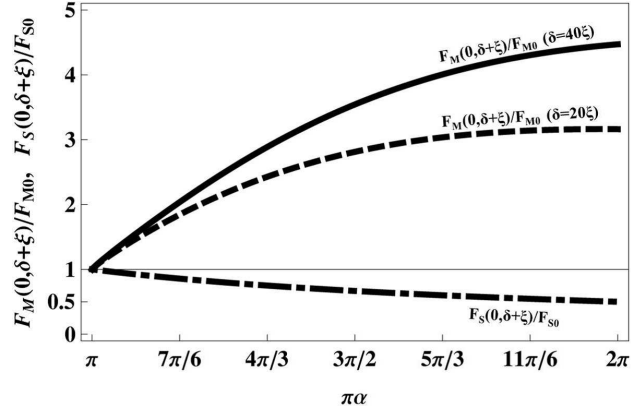}
   \end{center}\vspace{-0.2cm}
   \caption{
Forces acting on a vortex as a function of $\pi \alpha$. 
A solid curve and a dashed curve correpond to forces from an external magnetic field $F_{\rm M}(0, \delta+\xi)$ in an unit of $F_{\rm M0}$ with $\delta=40\xi$ and $20\xi$, respectively. 
A dashed-dotted curve corresponds to a force from the surface $F_{\rm S}(0, \delta+\xi)$ in an unit of $F_{\rm S0}$. 
   }\label{fig3}
\end{figure}

The force from the surface, ${\bf F}_{\rm S}$, can be expressed as ${\bf F}_{\rm S}={\bf J}_{\rm I} \times \phi_0 \hat{z}$, 
where ${\bf J}_{\rm I}$ is the image antivortex contribution to the total vortex current ${\bf J}_{\rm V+I}$. 
The calculation of ${\bf J}_{\rm V+I}$ can also be carried out by using the conformal mapping from the $w$-plane to the $\zeta$-plane (see Appendix~\ref{appendix_JVI}). 
The components of ${\bf J}_{\rm V+I}$ are given by~\cite{kuboLINAC14} 
\begin{eqnarray}
&& J_{{\rm V+I}x}(x,y) - i J_{{\rm V+I}y}(x,y) \nonumber \\
&=& \frac{1}{K_1 f(w)} \frac{-i\phi_0}{2\pi \mu_0 \lambda^2} \biggl( \frac{1}{w-i\epsilon} \!-\! \frac{1}{w+i\epsilon} \biggr)\biggr|_{w=F^{-1}(\zeta)} , 
\label{eq:JVI} 
\end{eqnarray}
where the first and the second term correspond to the vortex and image antivortex contributions, respectively.   
The total vortex current distribution calculated from Eq.~(\ref{eq:JVI}) is shown in Fig.~\ref{fig2}(b). 
Then we find $J_{{\rm I}x}(0,\delta+\xi) - i J_{{\rm I}y}(0,\delta+\xi) =  (i\phi_0/2\pi \mu_0 \lambda^2 K_1 f(i\epsilon)) (1/2i\epsilon) 
=\phi_0/4\pi \mu_0 \lambda^2 \xi\alpha$. 
Then the force from the surface is given by
\begin{eqnarray}
{\bf F}_{\rm S}(0,\delta+\xi)
= -\frac{\phi_0^2}{4\pi \mu_0 \lambda^2 \xi \alpha} \hat{\bf y} \,.  
\label{eq:FS}
\end{eqnarray}
which is identical with that given in the previous study on the groove with an infinite depth~\cite{buzdin}. 
The dependence on $\delta$ is dropped with the term $\mathcal{O}(\epsilon^2)$.  
Note that, when $\alpha\to 1$, Eq.~(\ref{eq:FS}) reproduces the force from the flat surface~\cite{kuboSRF2013, kuboIPAC14}, $F_{\rm S0}\equiv -\phi_0^2/4\pi \mu_0 \lambda^2 \xi$. 
In Fig.~\ref{fig3}, $F_{\rm S}$ in a unit of $F_{\rm S0}$ is shown as a function of $\pi \alpha$. 
As an angle increases, $F_{\rm S}$ decreases in contrast to $F_{\rm M}$.

%%%%%%%%%%%%%%%%%%%%%%%%%%%%%%%%%%%%%%%%%%%
\subsection{Suppression factor}
%%%%%%%%%%%%%%%%%%%%%%%%%%%%%%%%%%%%%%%%%%%

%
\begin{figure}[*t]
   \begin{center}
   \includegraphics[width=0.5\linewidth]{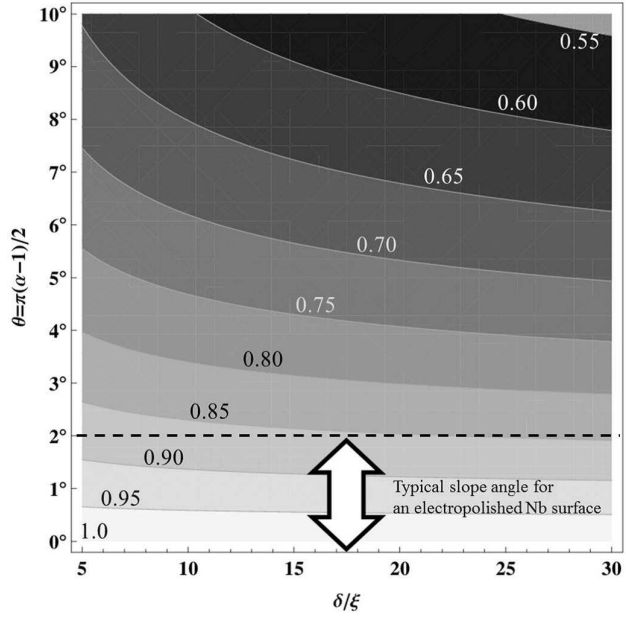}
   \end{center}\vspace{-0.2cm}
   \caption{
Contour plots of $\eta$. 
The abscissa represents the depth $\delta$ in a unit of $\xi$, and the ordinate represents the slope angle $\theta = \pi (\alpha-1)/2$. 
A region below the horizontal dashed line corresponds to typical slope-angles of the surface of electropolished Nb~\cite{xu}.
   }\label{fig4}
\end{figure}

$\widetilde{B}_s$ can be evaluated by balancing the two competing forces given by Eq.~(\ref{eq:FM}) and (\ref{eq:FS}). 
The surface current $J_0$ is given by $J_0 =B_0/ \mu_0 \lambda$, if the SC shown in Fig.~\ref{fig1} is the surface of semi-infinite SC. 
Then we find  
\begin{eqnarray}
\widetilde{B}_s  = \eta \, B_{s} \,, \hspace{1.5cm}  
\eta 
= \frac{1}{\alpha}
\biggl( \frac{\Gamma(\frac{\alpha}{2}) \Gamma(\frac{3-\alpha}{2}) \alpha  \sin\frac{\pi(\alpha-1)}{2}}{\sqrt{\pi}} \frac{\xi}{\delta}  \biggr)^{\!\!\!\frac{\alpha-1}{\alpha}} \!\!\!\!\!, 
\label{eq:Bs}
\end{eqnarray}
where $\eta$ is a suppression factor depending on a groove geometry, 
and $B_{s} \equiv B_c/\sqrt{2}\simeq 0.71\,B_c$ is the superheating field of the semi-infinite SC with the ideal flat surface in the London theory~\cite{bean, kuboSRF2013, kuboIPAC14}.
Fig.~\ref{fig4} shows a contour plot of $\eta$. 
As $\alpha$ and $\delta$ increase, $\eta$ decreases, 
because $F_{\rm M}$, which pushes a vortex into the inside, is increased, 
and $F_{\rm S}$, which prevents the vortex penetration, is decreased. 
It should be noted that, when a slope angle $\theta$ is small, $\eta$ is not sensitive to a defect depth, $\delta/\xi$.

A corresponding formula for the top SC layer of the multilayer coating can also be written in the same form as the above~\cite{kuboLINAC14},
$\widetilde{B}_s = \eta B_v^{\mathcal (S)}$, 
where $\eta$ is given by Eq.~(\ref{eq:Bs}), and $B_s$ is replaced by $B_v^{\mathcal (S)}$ given in the literature~\cite{kubo} (see Appendix~\ref{appendix_multilayer}).

%%%%%%%%%%%%%%%%%%%%%%
%%%%%%%%%%%%%%%%%%%%%%
\section{Discussion}
%%%%%%%%%%%%%%%%%%%%%%
%%%%%%%%%%%%%%%%%%%%%%

%
\begin{figure}[*t]
   \begin{center}
   \includegraphics[width=0.5\linewidth]{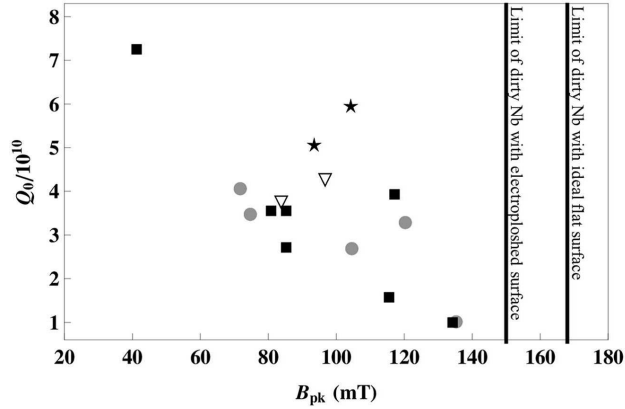}
   \end{center}\vspace{-0.2cm}
   \caption{
$B_{\rm pk}$ and $Q_0$ of nitrogen-doped Nb cavities at their achieved fields read from recent publications and presentations~\cite{crawford, romanenko, gonnellaLINAC14, furutaLCWS14, gengLCWS14, dhakal_ieee}. 
Squares represent results of cavities TE1AES016, TE1NR005, TE1AES003, TE1AES005, TE1AES013, TE1AES011, and TE1AES008 at $2\,{\rm K}$ by Fermilab~\cite{crawford, romanenko}, 
filled circles represent results of cavities LT1-1, LT1-2, LT1-3, LT1-4, and LT1-5 at $2\,{\rm K}$ by Cornell University~\cite{crawford, gonnellaLINAC14, furutaLCWS14}, 
stars represent results of cavity G2 at $1.8\,{\rm K}$ by Jeferson Lab~\cite{gengLCWS14}, 
and upside-down triangles represent results of cavities TD3 and TD4 at $2\,{\rm K}$ by Jeferson Lab~\cite{dhakal_ieee}. 
Theoretically evaluated field limits of dirty Nb with the electropolished surface ($1.5\times 10^2\,{\rm mT}$) and ideal flat surface  ($1.7\times 10^2\,{\rm mT}$)  are also shown. 
   }\label{fig5}
\end{figure}

By using Eq.~(\ref{eq:Bs}) and results of topographic studies, 
$\eta$ for surfaces of SCs with large GL parameters can be evaluated. 
As an immediate application, let us discuss $\widetilde{B}_s$ of dirty Nb processed by EP. 
Assuming surfaces of dirty Nb after EP have the same topography as the high-purity Nb processed by EP~\cite{xu},   
slope angles of surface topographies would distribute in $\lesssim 2^{\circ}$, 
which correspond to the area below the horizontal dashed line in Fig.~\ref{fig4}. 
In this region, $\eta$ is not sensitive to a defect size, $\delta/\xi$. 
Values just below the horizontal dashed line, 
\begin{eqnarray}
\eta \simeq 0.9 \,,
\end{eqnarray}
would define $\widetilde{B}_s$, 
at which vortex penetrations start at a large area of surface. 
Then we find 
\begin{eqnarray}\label{eq:BsEP1}
\widetilde{B}_{s}^{\rm (EP)} 
\simeq 0.9 \times 0.71 B_c \,,
\end{eqnarray}
where  $B_s \simeq 0.71 B_c$ of the London theory is used. 
Note here that $B_s \simeq 0.71 B_c$ is a good approximation at $T/T_c \simeq 1$ where it is close to $B_s \simeq 0.745 B_c$ of the GL or the quasi-classical (QC) theory~\cite{catelani, lin}, 
but is not necessarily a good approximation at a low temperature, $T/T_c \ll 1$. 
To evaluate $\widetilde{B}_s$ at $T/T_c \ll 1$ accurately, whole calculations should be carried out in the QC theory. 
We choose instead to improve the approximation by incorporating a correction based on the result of the QC theory: 
$B_{s}$ increases from $0.745 B_c$ at $T/T_c \simeq 1$ to $0.84 B_c$ at $T/T_c \ll 1$~\cite{catelani, lin}. 
Then we can estimate $\widetilde{B}_s^{\rm (EP)}$ at $T/T_c \ll 1$ as 
\begin{eqnarray}
\widetilde{B}_s^{\rm (EP)}\Bigr|_{\frac{T}{T_c} \ll1} 
\simeq 0.9 \times 0.84 B_c \,. 
\end{eqnarray}
Substituting $B_c(0\rm{K}) = 2.0\times 10^2\,{\rm mT}$, we obtain $\widetilde{B}_s^{\rm (EP)}|_{T/T_c\ll 1} \simeq 1.5\times 10^2 \,{\rm mT}$. 
Note that $\eta$ also depends on the temperature via $\xi$ and is proportional to $(\delta/\xi(T))^{-\frac{\alpha-1}{\alpha}}$, but is not sensitive to $\delta/\xi$ at a small slope angle as shown in Fig.~\ref{fig4}. 
Thus $\eta \simeq 0.9$ is thought to be valid at a broad temperature range. 
It is interesting to compare the above value with test results of nitrogen-doped Nb cavities.
As shown in Fig.~\ref{fig5}, $1.5\times 10^2 \,{\rm mT}$ is above the maximum field that has been achieved so far, and all other results are below it.  
In order to go beyond the limit of EP surface ($\eta \simeq 0.9$), a surface processing technology that can realize a further smooth surface with $\theta \ll 2^{\circ}$ is necessary. 
Mechanical polishing techniques that enable mirror-like finishes~\cite{cooper, palczewski, gengIPAC13} might be effective. 
On the other hand, for the case that the buffered chemical polishing (BCP) is applied instead of EP,  
surfaces have broader distributions of slope angle ($\lesssim 10^{\circ}$) as shown in the surface topographic study\cite{xu}, and $\eta$ would be further suppressed. 
In much the same way as the above, 
$\eta$ of surface of other materials with large GL parameters including multilayer SCs, and those after various surface processing technologies can also be evaluated by using Eq.~(\ref{eq:Bs}), if data of topographic studies are available.

It should be noted that a surface layer of Nb cavity after the low-temperature baking (LTB) is an example of the dirty-limit Nb, 
but the field limit of LTB-processed cavity can not be naively evaluated by using Eq.~(\ref{eq:Bs}),
because this system is not a simple semi-infinite SC. 
A penetration depth of LTB-processed Nb rapidly decreases in the first several tens of nm from the surface due to a depth-dependent mean free path~\cite{ciovati_bake, romanenko_bake}. 
This system may be modeled by layered SCs with different penetration depths. 
In such a system, it is known that a vortex is pushed to a direction of SC with a larger penetration depth~\cite{Mkrtchyan, kubo_bilayer}; 
a vortex is repelled from Nb with a smaller penetration depth behind the dirty layer. 
Thus, in a study of the field limit of LTB-processed Nb cavity, 
this non-trivial effect on the vortex dynamics should be carefully taken into account. 
This topic should also be addressed in a future work.

%%%%%%%%%%%%%%%%%%%%%%
%%%%%%%%%%%%%%%%%%%%%%
\section{Summary and outlook}
%%%%%%%%%%%%%%%%%%%%%%
%%%%%%%%%%%%%%%%%%%%%%

The field limit of SCRF cavity made of a type II SC with a large GL parameter has been studied with taking effects of nano-scale surface topography into account. 
We considered a triangular groove as a model of nano-defect and derived a formula for suppression factor of the superheating field in the framework of the London theory. 
Combining with a surface topographic study, a suppression factor of any surface of material can be evaluated. 
The formula was applied to the dirty Nb processed by EP as an example. 
The estimated field limit is consistent with the record field of nitrogen-doped Nb cavities. 
In much the same way as the eletropolished dirty Nb, 
suppression factors of surfaces of other bulk and multilayer superconductors, and those after various surface processing technologies can also be evaluated, 
which might explain what limits the field of these technologies.

In this paper, the formula of $\eta$ was derived in the framework of the London theory. 
For more comprehensive and accurate evaluations, 
whole calculations are needed to be self-consistently carried out by using the quasi-classical theory. 
Introducing a finite curvature radius of the corner might also be an interesting extension of this work.  
These works should be addressed in a future.

%%%%%%%%%%%%%%%%%%%%%%
%%%%%%%%%%%%%%%%%%%%%%
\section*{Acknowledgment}
%%%%%%%%%%%%%%%%%%%%%%
%%%%%%%%%%%%%%%%%%%%%%

The work is supported by JSPS Grant-in-Aid for Young Scientists (B), Number 26800157.

%%%%%%%%%%%%%%%%%%%%%%%%%%%
%%%%%%%%%%%%%%%%%%%%%%%%%%%

%%%%%%%%%
\appendix
%%%%%%%%%

%%%%%%%%%%%%%%%%%%%%%%%%%%%%%%%%%%%%%%%%%%%%%%%%%%%%%
%%%%%%%%%%%%%%%%%%%%%%%%%%%%%%%%%%%%%%%%%%%%%%%%%%%%%
\section{Screening current distribution} \label{appendix_JM}
%%%%%%%%%%%%%%%%%%%%%%%%%%%%%%%%%%%%%%%%%%%%%%%%%%%%%
%%%%%%%%%%%%%%%%%%%%%%%%%%%%%%%%%%%%%%%%%%%%%%%%%%%%%

${\bf J}_{\rm M}(x,y)$ can be derived by using the method of conformal mapping as follows~\cite{kuboLINAC14}. 
${\bf J}_{\rm M}$ satisfies ${\rm div}\,{\bf J}_{\rm M}=0$ and one of the Maxwell equations ${\bf J}_{\rm M}={\rm rot}\,{\bf H}$, 
where the magnetic field ${\bf H}$ plays the role of the vector potential of ${\bf J}_{\rm M}$. 
For our setup, $\bf H$ can be written as ${\bf H}=(0, 0, $ $-\psi(x,y))$, 
and ${\bf J}_{\rm M}$ is given by ${\bf J}_{\rm M} ={\rm rot}\,{\bf H}=(-\partial \psi/\partial y, \partial \psi/\partial x, 0)$. 
On the other hand, since $\lambda$ is assumed to be much larger than the typical scale of the model, $\delta$, 
the London equation is reduced to ${\rm rot}\,{\bf J}_{\rm M} = -\triangle {\bf H}= {\bf 0}$, 
which allows us to introduce a scalar potential of ${\bf J}_{\rm M}$. 
For our setup, the scalar potential can be written as $\phi(x,y)$, 
and ${\bf J}_{\rm M}$ is given by ${\bf J}_{\rm M} = -{\rm grad}\,\phi  = (-\partial \phi/\partial x, -\partial \phi/\partial y, 0)$. 
Since both the two approaches should lead the same ${\bf J}_{\rm M}$, we find
\begin{eqnarray}
J_{{\rm M}x} = -\frac{\partial \phi}{\partial x} = -\frac{\partial \psi}{\partial y} \,, \hspace{1cm} 
J_{{\rm M}y} = -\frac{\partial \phi}{\partial y} = \frac{\partial \psi}{\partial x} \,,
\end{eqnarray}
which are the Cauchy-Riemann conditions.  
Thus a function defined by
\begin{eqnarray}
\Phi_{\rm M}(\zeta) \equiv \phi(x,y) + i\psi(x,y)\,,
\end{eqnarray}
is an holomorphic function of a complex variable $\zeta=x+iy$, which is called the complex potential.  
If $\Phi_{\rm M}(\zeta)$ is given, components of ${\bf J}_{\rm M}$ are derived from 
\begin{eqnarray}
J_{{\rm M}x}- i J_{{\rm M}y} 
= - \frac{\partial \phi}{\partial x} - i \biggl( -\frac{\partial \phi}{\partial y} \biggr)
= -\frac{\partial \phi}{\partial x} - i\frac{\partial \psi}{\partial x} 
= -\frac{d\Phi_{\rm M}(\zeta)}{d\zeta}  \,, 
\label{eqA3}
\end{eqnarray}
where the property of the holomorphic function, $\Phi_{\rm M}'(\zeta)=\partial \phi/\partial x + i\partial \psi/\partial x$, is used. 
Then our problem is reduced to that of finding $\Phi_{\rm M}(\zeta)$.

$\Phi_{\rm M}(\zeta)$ can be derived from that on the $w$-plane, $\widetilde{\Phi}_{\rm M}(w)$, through a conformal mapping $\zeta =F(w)$. 
The map is given by the Schwarz-Christoffel transformation,
\begin{eqnarray}
\zeta =F(w) = K_1 \int_0^w \!\! f(w')  dw' + K_2 \, , \label{eqA4}
\end{eqnarray}
where $f(w)$ is given by
\begin{eqnarray}
f(w) = w^{\alpha -1}(w^2 - 1)^{-\frac{\alpha-1}{2}} \,,
\end{eqnarray}
and $K_1$ and $K_2$ are constants determined by the conditions that A' and B' on the $w$-plane are mapped into A and B on the $\zeta$-plane, respectively. 
$\widetilde{\Phi}_{\rm M}(w)$ is given by 
\begin{eqnarray}
\widetilde{\Phi}_{\rm M}(w) = K_1 J_0 w
\end{eqnarray}
which yields the current distribution on the $w$-plane by a similar equation as Eq.~(\ref{eqA3}), 
$\widetilde{J}_{{\rm M}u}(u,v)- i \widetilde{J}_{{\rm M}v}(u,v) =- d\widetilde{\Phi}_{\rm M}(w)/dw = -K_1 J_0 \equiv -\widetilde{J}_0$. 
Then 
\begin{eqnarray}
\Phi_{\rm M}(\zeta) = \widetilde{\Phi}_{\rm M}(w)\bigr|_{w=F^{-1}(\zeta)} = K_1 J_0 F^{-1}(\zeta) \,,
\end{eqnarray}
where $F^{-1}$ is an inverse function of $F$. 
Then Eq.~(\ref{eqA3}) becomes
\begin{eqnarray}
J_{{\rm M}x}(x,y) - i J_{{\rm M}y}(x,y) 
= -\frac{d\Phi_{\rm M}(\zeta)}{d\zeta}
= -\frac{K_1 J_0}{dF/dw}\biggr|_{w=F^{-1}(\zeta)}
= -\frac{J_0}{f(w)}\biggr|_{w=F^{-1}(\zeta)}    \,,
\label{eqA8}
\end{eqnarray}
where $dF^{-1}/d\zeta = dw/d\zeta= (d\zeta/dw)^{-1} = (dF/dw)^{-1}$ is used. 
Note that Eq.~(\ref{eqA8}) reproduces the current density far from the groove, $-J_0$, when $z\to\infty$ or $w\to \infty$.

\begin{figure}[*t]
   \begin{center}
   \includegraphics[width=0.5\linewidth]{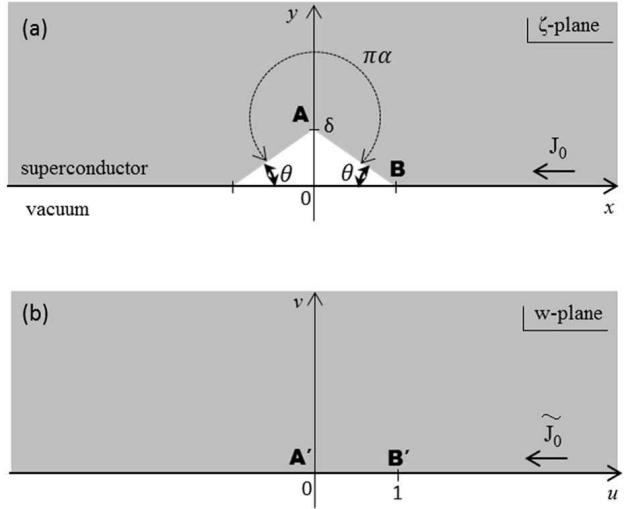}
   \end{center}\vspace{-0.2cm}
   \caption{
(a) Triangular groove on the $\zeta$-plane and 
(b) its map on the $w$-plane. 
   }\label{figA}
\end{figure}
%

%%%%%%%%%%%%%%%%%%%%%%%%%%%%%%%%%%%%%%%%%%%%%%%%%%%%%
%%%%%%%%%%%%%%%%%%%%%%%%%%%%%%%%%%%%%%%%%%%%%%%%%%%%%
\section{Explicit forms of $K_1$ and $K_2$} \label{appendix_K1K2}
%%%%%%%%%%%%%%%%%%%%%%%%%%%%%%%%%%%%%%%%%%%%%%%%%%%%%
%%%%%%%%%%%%%%%%%%%%%%%%%%%%%%%%%%%%%%%%%%%%%%%%%%%%%

Explicit forms of $K_1$ and $K_2$ are obtained by imposing the conditions (i) $\rm{A}'$ is mapped into $\rm{A}$ and (ii) $\rm{B}'$ into $\rm{B}$~\cite{kuboLINAC14}. 
Imposing the condition (i) on Eq.~(\ref{eqA4}), we find
\begin{eqnarray}
i\delta = K_1 \int_0^0 \!\! f(w')  dw' + K_2 = K_2 \,. 
\end{eqnarray}
Similarly, imposing the condition (ii) on Eq.~(\ref{eqA4}), we find
\begin{eqnarray}
\frac{\delta}{\tan \theta} = K_1 \int_0^1 \!\!\! dw\, w^{\alpha -1}(w^2 - 1)^{-\frac{\alpha-1}{2}} + i\delta \, , 
\end{eqnarray}
Since $\theta=\pi(\alpha-1)/2$, the above condition becomes 
\begin{eqnarray}
\frac{e^{-i \frac{\pi(\alpha-1)}{2}} \delta}{K_1 \sin\frac{\pi(\alpha-1)}{2}} 
&=& \int_0^1 \!\!\! dw\, w^{\alpha -1}(w^2 - 1)^{-\frac{\alpha-1}{2}}  \nonumber \\
&=& e^{-i \frac{\pi(\alpha-1)}{2}} \int_0^1 \!\!\! dw\, w^{\alpha -1}(1-w^2)^{-\frac{\alpha-1}{2}} 
\, . 
\end{eqnarray}
Replacing $w$ with $t\equiv w^2$, we find
\begin{eqnarray}
\frac{\delta}{K_1 \sin\frac{\pi(\alpha-1)}{2}} 
&=& \frac{1}{2} \int_0^1 \!\!\! dt\, t^{\frac{\alpha}{2} -1} (1-t)^{\frac{3-\alpha}{2}-1}  \nonumber \\
&=& \frac{1}{2} \frac{\Gamma(\frac{\alpha}{2})\Gamma(\frac{3-\alpha}{2})}{\Gamma(\frac{3}{2})} \, .
\end{eqnarray}
Then we finally obtain 
\begin{eqnarray}
K_1 = \frac{\sqrt{\pi} \delta}{\Gamma(\frac{\alpha}{2})\Gamma(\frac{3-\alpha}{2}) \sin\frac{\pi(\alpha-1)}{2}} 
\, . 
\end{eqnarray}
%

%%%%%%%%%%%%%%%%%%%%%%%%%%%%%%%%%%%%%%%%%%%%%%%%%%%%%
%%%%%%%%%%%%%%%%%%%%%%%%%%%%%%%%%%%%%%%%%%%%%%%%%%%%%
\section{Vortex current distribution} \label{appendix_JVI}
%%%%%%%%%%%%%%%%%%%%%%%%%%%%%%%%%%%%%%%%%%%%%%%%%%%%%
%%%%%%%%%%%%%%%%%%%%%%%%%%%%%%%%%%%%%%%%%%%%%%%%%%%%%

${\bf J}_{\rm V+I}(x,y)$ can be derived as follows~\cite{kuboLINAC14}. 
A current associated with a vortex near a surface satisfies the boundary condition of zero current normal to the surface. 
Such a current distribution can be reproduced by removing the surface and introducing an appropriate image antivortex. 
Then the total vortex current is given by a summation of currents due to a vortex and an image antivortex on an infinite SC without a surface.  
Since the vortex and the image antivortex on the $w$-plane are located at $w=+i\epsilon \simeq i (\alpha \xi/K_1)^{\frac{1}{\alpha}}$ and $-i\epsilon$, respectively, 
the total vortex current distribution on the $w$-plane, $\widetilde{\bf J}_{\rm V+I}$, is given by 
\begin{eqnarray}
\widetilde{J}_{{\rm V+I}u}(u,v)- i \widetilde{J}_{{\rm V+I}v}(u,v)  
=\frac{i \phi_0}{2\pi \mu_0 \lambda^2} \biggl( \frac{1}{w-i\epsilon}  -\frac{1}{w+i\epsilon} \biggr)
\,, 
\end{eqnarray}
and the complex potential on the $w$-plane, $\widetilde{\Phi}_{\rm V+I}(w)$, is given by
\begin{eqnarray}
\widetilde{\Phi}_{\rm V+I}(w) 
= \frac{i \phi_0}{2\pi \mu_0 \lambda^2} \bigl[\log (w-i\epsilon)  -\log (w+i\epsilon)\bigr] \,.  
\end{eqnarray}
Then the complex potential on the $\zeta$-plane, $\Phi_{\rm V+I}(\zeta)$, is given by 
\begin{eqnarray}
\Phi_{\rm V+I}(\zeta)=\widetilde{\Phi}_{\rm V+I}(w)\biggr|_{w=F^{-1}(\zeta)} \,,
\end{eqnarray}
and the toatal vortex current distribution on the $\zeta$-plane, ${\bf J}_{\rm V+I}$, is given by 
\begin{eqnarray}
J_{{\rm V+I}x}(x,y) - i J_{{\rm V+I}y}(x,y) 
&=& -\frac{d\Phi_{\rm V+I}(\zeta)}{d\zeta}
= -\frac{1}{dF/dw}\frac{d\widetilde{\Phi}_{\rm V+I}(w)}{dw}\biggr|_{w=F^{-1}(\zeta)} \nonumber \\
&=& \frac{1}{K_1 f(w)} \frac{-i\phi_0}{2\pi \mu_0 \lambda^2} \biggl( \frac{1}{w-i\epsilon} \!-\! \frac{1}{w+i\epsilon} \biggr)\biggr|_{w=F^{-1}(\zeta)}  . 
\end{eqnarray}
where the explicit form of $K_1$ is given in the last section. 

%%%%%%%%%%%%%%%%%%%%%%%%%%%%%%%%%%%%%%%%%%%%%%%%%%%%%
%%%%%%%%%%%%%%%%%%%%%%%%%%%%%%%%%%%%%%%%%%%%%%%%%%%%%
\section{Suppression factor for the multilayer coating} \label{appendix_multilayer}
%%%%%%%%%%%%%%%%%%%%%%%%%%%%%%%%%%%%%%%%%%%%%%%%%%%%%
%%%%%%%%%%%%%%%%%%%%%%%%%%%%%%%%%%%%%%%%%%%%%%%%%%%%%

The suppression factor for the multilayer coating can be derived in much the same way as that for the semi-infinite SC~\cite{kuboLINAC14}. 
When the SC shown in Fig.~\ref{fig1} is a part of the top SC layer of the multilayer coating, 
the surface current $J_0$ should be calculated by using the correct magnetic-field distribution~\cite{kubo} and is given by 
\begin{eqnarray}
J_0 = \frac{B_0}{\mu_0 \lambda}
\frac{ \sinh \frac{d_{\mathcal{S}}}{\lambda} + (\frac{\lambda'}{\lambda} + \frac{d_{\mathcal{I}}}{\lambda}) \cosh \frac{d_{\mathcal{S}}}{\lambda}}
     { \cosh \frac{d_{\mathcal{S}}}{\lambda} + (\frac{\lambda'}{\lambda} + \frac{d_{\mathcal{I}}}{\lambda}) \sinh \frac{d_{\mathcal{S}}}{\lambda}} \, , 
\label{eq:D1}
\end{eqnarray}
where $d_{\mathcal S}$ is a thickness of the top SC layer, $d_{\mathcal I}$ is a thickness of the insulator layer, and $\lambda'$ is a penetration depth of the SC substrate. 
Substituting Eq.~(\ref{eq:D1}) into Eq.~(\ref{eq:FM}) and balancing Eq.~(\ref{eq:FM}) and (\ref{eq:FS}), we obtain
\begin{eqnarray}
\widetilde{B}_s = \eta \,B_v^{\mathcal(S)} \,,
\end{eqnarray}
where 
\begin{eqnarray}
B_v^{\mathcal(S)}
=
\frac{\cosh\frac{d_{\mathcal{S}}}{\lambda} + (\frac{\lambda'}{\lambda} + \frac{d_{\mathcal{I}}}{\lambda})\sinh\frac{d_{\mathcal{S}}}{\lambda}}
     {\sinh\frac{d_{\mathcal{S}}}{\lambda} + (\frac{\lambda'}{\lambda} + \frac{d_{\mathcal{I}}}{\lambda})\cosh\frac{d_{\mathcal{S}}}{\lambda} } B_s 
 \,, 
\end{eqnarray}
is the enhanced superheating field of the top SC layer with an ideal flat surface~\cite{kubo}.

\end{document}